\documentclass[aps,prl,twocolumn,superscriptaddress,showpacs]{revtex4}
\usepackage{amsmath,amssymb,bm}
\usepackage{graphicx,epsfig}
\renewcommand{\v}[1]{{\bf #1}}
\newcommand{\be}{\begin{equation}}
\newcommand{\ee}{\end{equation}}
\newcommand{\nn}{\nonumber \\}
\newcommand{\ij}{\langle ij \rangle}

\newcommand{\ba}{\begin{eqnarray}}
\newcommand{\ea}{\end{eqnarray}}

\newcommand{\bw}{\begin{widetext}}
\newcommand{\ew}{\end{widetext}}

\newcommand{\bpm}{\begin{pmatrix}}
\newcommand{\epm}{\end{pmatrix}}

\begin{document}

\title{Nematic and chiral orders for planar spins on triangular lattice}

%
\author{Jin-Hong Park}
\affiliation{Department of Physics, BK21 Physics Research Division,
Sungkyunkwan University, Suwon 440-746, Korea}
\author{Shigeki Onoda}
%
\affiliation{Condensed Matter Theory Laboratory, RIKEN, 2-1,
Hirosawa, Wako 351-0198, Japan}
\author{Naoto Nagaosa}
\affiliation{Department of Applied Physics, The University of Tokyo,
7-3-1, Hongo, Bunkyo-ku, Tokyo 113-8656, Japan} \affiliation{Cross
Correlated Materials Research Group, Frontier Research System,
Riken, 2-1 Hirosawa, Wako, Saitama 351-0198, Japan}
\author{Jung Hoon Han}
\email[Corresponding author: $~~$]{hanjh@skku.edu}
\affiliation{Department of Physics, BK21 Physics Research Division,
Sungkyunkwan University, Suwon 440-746, Korea} 
\begin{abstract} We propose a variant of the antiferromagnetic XY model on
the triangular lattice to study the interplay between the chiral and
nematic orders in addition to the magnetic order. The model has a
significant bi-quadratic interaction of the planar spins. When the
bi-quadratic exchange energy dominates, a large temperature window is
shown to exist over which the nematic and the chiral orders co-exist
without the magnetic order, thus defining a chiral-nematic state. The
phase diagram of the model and some of its critical properties are
derived by means of the Monte Carlo simulation.
\end{abstract}

\maketitle

Nontrivial orders in frustrated magnets~\cite{Villain} are among the
central issues in the field of condensed-matter physics. Besides the
conventional magnetic order parameter of spin $\v S_i$ at a site $i$,
there could appear various nontrivial orders such as
vector~\cite{MiyashitaShiba84,KawamuraTanemura87} and
scalar~\cite{Baskaran89,WenWilczekZee89} chiral
orders~\cite{Richter92}, and nematic order~\cite{Chalker92}, which
might lead to additional phase transitions distinct from the one
driven by magnetic order. Even the ground state itself may be
characterized solely by these nontrivial orders. This issue is now
attracting revived interest from the viewpoint of nontrivial glass
transition of spins~\cite{KawamuraLi01} and multiferroic
behaviors~\cite{OnodaNagaosa07,FSSO08}, where the ferroelectricity is
induced by the formation of vector spin chirality~\cite{mf}. One
important aspect of this problem is the interplay between the various
orders. Usually the nontrivial orders become long ranged when the
magnetic order sets in. For example, the spiral spin order naturally
implies the vector spin chiral order through $\langle {\v S_i} \times
{\v S_j}\rangle \sim \langle \v S_i \rangle \times \langle \v S_j
\rangle$ on the neighboring sites. Therefore, the interesting issue
is whether the nontrivial order can become long ranged \textit{in the
absence of} the magnetic order. This issue has been studied
theoretically~\cite{OnodaNagaosa07}, and experimentally in the
quasi-one dimensional frustrated magnet~\cite{Cinti08} where the
chiral order appears above the magnetic phase transition. The next
important question, we argue, is the interplay between the two
nontrivial orders, e.g., chiral and nematic orders, which has not
been fully addressed so far.

To address this issue, we study a generalized classical XY spin model
on a triangular lattice,

\be H = J_1 \sum_{\ij} \cos ( \theta_{ij} )+ J_2 \sum_{\ij} \cos (
2\theta_{ij} ) , \label{eq:our-model}\ee
where $\theta_{ij}$ is the angle difference $\theta_i -\theta_j$
between the nearest neighbors $\ij$. This model contains the usual
frustration in the exchange interaction due to the triangular lattice
geometry, together with the possible nematic order induced by the
$J_2$ term. The $J_2=0$ limit has been extensively studied, and it is
believed to have two phase transitions at closely spaced critical
temperatures\cite{MiyashitaShiba84,olsson,leelee,new-literature}. The
Kosterlitz-Thouless (KT) transition temperature $T_\mathrm{KT}$
signaling the loss of (algebraic) magnetic order and the melting
temperature of the staggered chirality, $T_\chi$, are extremely
close,  $( T_\chi - T_\mathrm{KT}) /T_\chi \lesssim 0.02$ at $J_2=0$,
hampering the interpretation of the intermediate, $T_\mathrm{KT} < T
< T_\chi$ phase as the chiral phase in which the chirality is ordered
but the magnetism remains disordered. Extension of the XY model to
include large $J_2$ interaction was considered earlier in Refs.
\cite{dhlee,chalker}, where the authors examined the phase diagram of
Eq. (\ref{eq:our-model}) \textit{on the square lattice}, which lacks
frustration. In contrast, our model on the triangular lattice serves
as a minimal model to study the two nontrivial orders, i.e., the
chiral order induced by the geometric frustration, and the nematic
order induced by the bi-quadratic interaction.

A unique feature of the large $J_2/J_1$ region of the model as noted
in Refs. \cite{dhlee,chalker} is the existence of an Ising phase
transition associated with the vanishing string tension between
half-integer vortices in addition to the KT transition. This Ising
phase transition turns out to correspond to the onset of the
(algebraic) magnetic order. Being driven by $J_1$, the Ising
transition temperature occurs at a much lower temperature than either
the chiral or the nematic transition, which are both driven by $J_2$.
The result is the existence of a magnetism-free, chiral-nematic phase
in the large $J_2 /J_1$ part of our model.

\begin{figure}[t]
\centering
\includegraphics[scale=0.35]{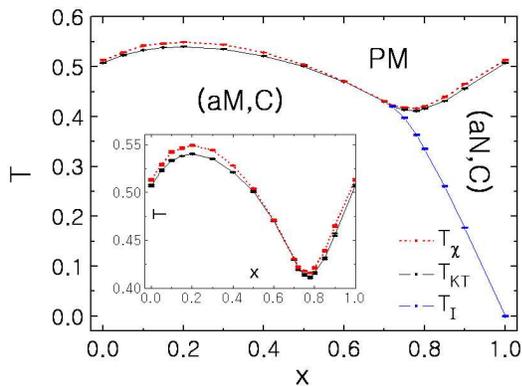}
\caption{(Color online) Phase diagram of the $J_1 - J_2$ model in Eq.
(\ref{eq:our-model}) with $J_1 = 1-x$ and $J_2 = x$. Two closely
spaced transition temperatures labeled by $T_\mathrm{KT}$ and
$T_\chi$ separate the paramagnetic (PM) phase from the algebraically
correlated phase at a lower temperature. aM, aN, and C stand for
phases with algebraic correlations in (antiferro)magnetic and
(antiferro)nematic order parameters, and the long-range correlations
in the chirality order. A further transition from aN to aM occurs as
an Ising transition for $x>x_c$ with $x_c \approx 0.7$. All the
symbols have a thickness in the temperature direction consistent with
their statistical errors. (inset) The onset of chirality order at
$T_\chi$ (red) takes place at temperatures close to, but slightly
higher than the corresponding KT transition temperature
$T_\mathrm{KT}$ (black) for all $x$. The Ising transition temperature
$T_I$ is not shown here for clarity.} \label{fig:phase-diagram}
\end{figure}

\textbf{Phase diagram}:  The $x\!-\!T$ phase diagram for Eq.
(\ref{eq:our-model}) is shown in Fig. \ref{fig:phase-diagram}, where
$T$ is the temperature and $x$ parameterizes the interaction as $J_1
= 1-x$, $J_2 = x$. Detailed Monte Carlo (MC) calculations were
performed with $5\times 10^5$ MC steps per run, on $L\times L$
lattice with $L$ ranging from 15 to 60. Occasional checks were made
on a larger lattice of up to $L=100$ to ensure that no discernible
changes in either the critical temperatures or the critical exponents
are obtained from the larger size. Typically, $10^5$ steps were
discarded to reach equilibrium. An integer vortex-mediated KT
transition marking the PM-aM boundary bifurcates into a half-integer
vortex-mediated KT transition, marking the PM-aN boundary, plus an
Ising transition\cite{dhlee} when $x$ exceeds $x_c\approx 0.7$. The
Ising transition in turn separates the aM from aN. For the whole
range of $x$, the chiral transition temperature $T_\chi$ stays
slightly above $T_\mathrm{KT}$, with the possible exception at
$x=x_c$ where they may coincide.
\\

\begin{figure}[t]
\centering
\includegraphics[scale=0.4]{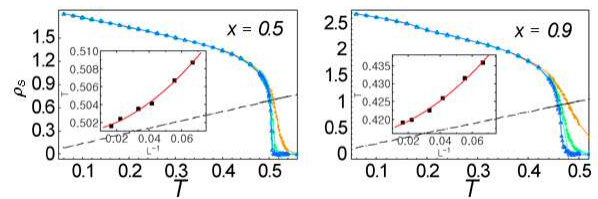}
\caption{(Color online) Phase stiffness $\rho_s (T)$ according to
Eq. (\ref{rho}) for $L=15-60$ and $x=0.5$ and $0.9$. The straight
line is $ (2 / \pi)( \sqrt{3}/ 2)(J_1+4J_2)T$. The crossing
temperature of this line and $\rho_s (T)$ for each $L$ is shown in
the inset along with the extrapolation to $L^{-1}=0$.}
\label{fig:helicity-modulus}
\end{figure}

\textbf{KT transition at $T_\mathrm{KT}$}: The determination of
$T_\mathrm{KT}$ is made with the phase stiffness, also called the
helicity modulus, appropriate for the $J_1 \!-\! J_2$ model

\ba && \rho_s (T) =-{J_1 \over 2 L^2} \langle \sum_{\langle ij
\rangle} \cos \theta_{ij} \rangle  -{2 J_2 \over L^2}\langle \sum_{
\langle ij \rangle} \cos 2 \theta_{ij} \rangle \nn &&
- {1 \over T L^2} \langle ( J_1 \sum_{\langle ij \rangle} x_{ij}\sin
\theta_{ij} +2 J_2 \sum_{\langle ij \rangle} x_{ij} \sin 2
\theta_{ij} )^{2} \rangle .\ \ \ \ \  
 \label{rho}\ea
Here $x_{ij} = x_i - x_j$ is the separation of the $x$-coordinate.
The crossing of $\rho_s (T)$ with the straight line $(2 /\pi)(
\sqrt{3}/ 2 ) (J_1 + 4J_2 )T=(2 /\pi)( \sqrt{3}/ 2 ) (1+3x)T$ yields,
for a given lattice size $L$, an estimate of the critical temperature
$T_\mathrm{KT} (L)$\cite{leelee}. Extrapolation to $L\rightarrow
\infty $ using polynomial fits as shown in the insets of Fig.
\ref{fig:helicity-modulus} yields the estimate of $T_\mathrm{KT}$. A
more sophisticated method taking into account the logarithmic
correction\cite{weber} yields a similar answer\cite{new-literature}.
\\

\textbf{Chirality transition at $T_\chi$}: It is customary to define
the chirality $\chi$ as the directed sum of the bond current
$\langle \sin \theta_{ij} \rangle$\cite{olsson} following the
relation $\langle \sin\theta_{ij} \rangle \sim - \partial F/\partial
A_{ij}$. The free energy $F$ is evaluated with respect to the
modified interaction $\cos \theta_{ij} \rightarrow \cos
(\theta_{ij}+ A_{ij})$. A similar modification of Eq.
(\ref{eq:our-model}) results in the bond current

\be J_{ij} \sim J_1 \langle \sin (\theta_{ij}) \rangle + 2J_2
\langle \sin (2\theta_{ij}) \rangle .\label{eq:new-chirality} \ee
This new definition is particularly effective as $x \rightarrow 1$,
where the conventional definition $\sim \langle \sin \theta_{ij}
\rangle $ vanishes identically due to the $Z_2$ symmetry. For each
$x$, $T_\chi$ was obtained from Binder cumulant analysis for the new
definition of chirality based on Eq. (\ref{eq:new-chirality}). The
conventional definition $(J_2 =0$) gave an estimate of $T_\chi$
which differs only in the third significant digit. Although our
analysis showed $T_\chi \gtrsim T_\mathrm{KT}$ for all $x$, we do
not at present rule out the scenario in which $T_\chi$ and
$T_\mathrm{KT}$ merge at $x=x_c$, resulting in a multi-critical
point there. If that happens, the second-order chirality transition
may become weakly first-order.

\begin{figure}[t]
\centering
\includegraphics[scale=0.55]{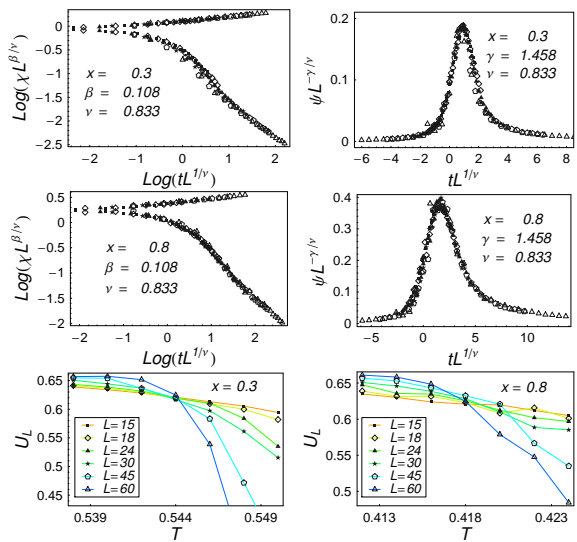}
\caption{(Color online) A scaling plot of chirality based on Eq.
\ref{eq:new-chirality}) and its susceptibility for $x=0.3$ and
$x=0.8$, for lattice sizes $L=15-60$. The exponents used are those
of $x=0$\cite{leelee}. The last row shows the behavior of the Binder
cumulants at $x=0.3$ and $x=0.8$, respectively.}
\label{fig:chi-scaling}
\end{figure}

Earlier analysis\cite{leelee} at $x=0$ identified the transition of
$\chi$ with the non-Ising critical exponents $1/\nu =1.2$, and
$\beta/\nu=0.12$, $\gamma/\nu=1.75$. Figure \ref{fig:chi-scaling}
shows $\chi$ and its variant, $\psi \equiv (\langle \chi^2 \rangle -
\langle \chi \rangle^2 )/T$, in scaling form $ \mathcal{\chi}
=L^{-\beta /\nu}  f \left( t L^{1/\nu}\right)$, $\psi  =L^{\gamma
/\nu} g \left( t L^{1/\nu}\right)$, with $t=|T-T_\chi |/T_\chi$, at
$x=0.3$ and $x=0.8$. Same exponents as for the $x=0$ case works well
in scaling throughout the whole phase diagram. Appearance of the
non-Ising exponents for $J_2=0$ have been explained in terms of an
enhanced finite-size scaling effect at small sizes due to the
screening length associated with the KT transition, in the cases of
the square lattice\cite{olsson} and triangular lattice\cite{leelee}.
Here it is equally possible that the true universality classes at
$T_\chi$ that of Ising transition. At any rate, the identification
of the chirality transition $T_\chi$ well above the magnetic
transition for large $J_2 /J_1$ ratio is unequivocal and proves the
existence of the magnetism-free, chiral-nematic phase in our model.
\\

\begin{figure}[t]
\centering
\includegraphics[scale=0.25]{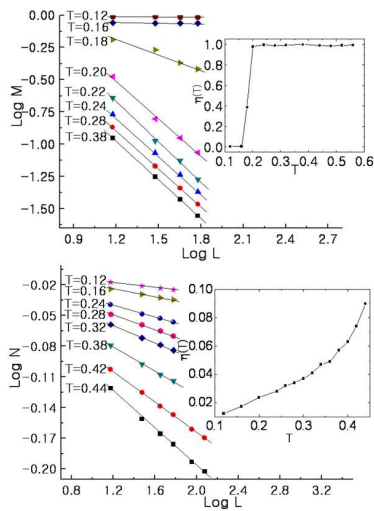}
\caption{(Color online) The size dependence of (a) the magnetic
($\mathcal{M}$) and (b) the nematic ($\mathcal{N}$) order parameters
at $x=0.9$ are shown on the log-log plot. (insets) The critical
exponent $\eta_{\mathcal{M}} (T)$ for $\mathcal{M}\sim
1/L^{\eta_{\cal{M}} (T)}$ and $\eta_{\cal{N}} (T)$ for
$\mathcal{N}\sim 1/L^{\eta_{\cal{N}} (T)}$.}
\label{fig:MN-data}
\end{figure}

\textbf{Magnetic and nematic orders}: The low-temperature phase
immediately below $T_\mathrm{KT}$ is either aM or aN, depending on
whether $x<x_c$ or $x>x_c$. The magnetic and nematic correlations
are examined on the basis of the order parameters, $\mathcal{M} =
(3/L^2 ) | \sum_{i\in \mathcal{A}} e^{i\theta_i} |$, and
$\mathcal{N} = (3/L^2 ) | \sum_{i\in \mathcal{A}} e^{2i\theta_i} |$,
respectively, where the sum $i\in \mathcal{A}$ spans the
$\mathcal{A}$ sublattice sites. For $T_\mathrm{I}<T<T_\mathrm{KT}$,
the magnetic order parameter is expected to lose its algebraic
character and become short-ranged. Indeed, the size dependence of
$\mathcal{M}$ as revealed by $\mathcal{M}\sim 1/L^{\eta_{\cal{M}}
(T)}$ for $x=0.9$ has the exponents $\eta_{\cal{M}} (T)$ changing
abruptly from $\approx 1$ above $T_\mathrm{I}$ to a small value
below it (Fig. \ref{fig:MN-data} (a)). The critical nature of the
nematic order parameter $\mathcal{N}$ at $x >x_c $ is seen in the
continuous dependence of the exponent $\eta_{\cal{N}}$,
$\mathcal{N}\sim 1/L^{\eta_{\cal{N}} (T)}$, as shown in Fig.
\ref{fig:MN-data} (b) for $x=0.9$. The $T$-dependent exponent
$\eta_{\cal{N}} (T)$ continuously decreases as the temperature is
lowered, even in the low-$T$ magnetic phase $T<T_\mathrm{I}$,
indicating that the nematic order remains critical in the whole
temperature range $0< T < T_\mathrm{KT}$. A careful comparison of
$\eta_{\cal{M}} (T)$ and $\eta_{\cal{N}}(T)$ for $T$ below
$T_\mathrm{I}$  revealed a relation $\eta_{\cal{N}}(T) \approx 4
\eta_{\cal{M}}(T)$, in accord with the expectations of the spin wave
analysis.
\\

\begin{figure}[t]
\centering
\includegraphics[scale=0.3]{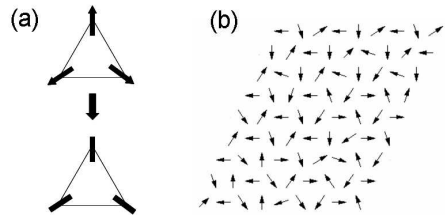}
\caption{(a) A cartoon depicting the loss of head-tail order in going
from aM to aN phase. (b) A snapshop of the chiral-nematic state at
$T=0.2$ for $x=0.9$ where the Ising transition occurs at $T_I =
0.177$. Within the same sublattice the ``body" of the arrows, not
their tips, are seen to point in the same general direction. }
\label{fig:chiral-nematic}
\end{figure}


\textbf{Ising transition at $T_\mathrm{I}$}: A cartoon picture of the
Ising transition is given in Fig. \ref{fig:chiral-nematic} (a), where
it is described as the loss of local ``head-tail" order. The choice
of the order parameter for the transition is not unique and, to the
best of our knowledge, has never been given an explicit expression.
Here we choose to analyze the temperature dependence of

\be \mathcal{I}=(3/L^2 ) \sum_{i\in \mathcal{A}}
\mathrm{sgn}(\cos[\theta_i -\theta_{i0}]) , \label{eq:Ising-OP}\ee
where $\theta_{i0}$ is the spin angle at some reference site $i0$ of
the $\mathcal{A}$ sublattice. As an Ising-like variable,
$\mathrm{sgn}(\cos[\theta_i -\theta_{i0}])$ carries two allowed
values $\pm 1$. In the aN phase, $\theta_i$ and $\theta_i +\pi$ occur
with equal probabilities, thus $\mathcal{I}=0$. An excellent data
collapse in finite-size scaling was obtained with the 2D Ising
critical exponents, $\beta=1/8$, $\gamma=1.75$, and $\nu =1$ for both
$x=0.8$ and $x=0.9$. To be exact, the orientation of $\theta_i$ with
regard to a reference angle $\theta_{i0}$ will be arbitrary as the
separation $i-i0$ tends to infinity in a truly thermodynamic system.
Given the small exponent $\eta_{\cal N} (T) < 0.03$ near $T=T_I$
consistent with an extremely slow decay, however, one can argue that
the only effective low-energy fluctuation is the $\pi$-flip of the
spin (which reverses the sign of $\cos [\theta_i \!-\!\theta_{i0}]$)
rather than the small-angle fluctuations (which does not reverse the
sign) for the practical system sizes considered in the MC simulation.
As far as this is the case, our definition serves as a good measure
of the Ising transition.
\\
%

\begin{figure}[ht]
\centering
\includegraphics[scale=0.3]{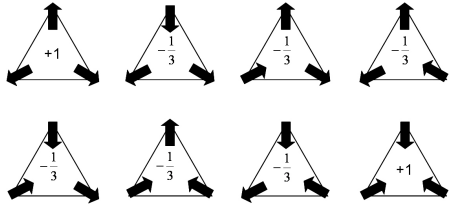}
\caption{Eight possible magnetic patterns within the nematically
ordered phase, which also includes configurations with the global
rotation of all the spins shown here. The corresponding chirality of
each spin configuration is shown inside the triangle.}
\label{fig:triangle}
\end{figure}

\textbf{Chiral-nematic phase}: The central finding of this work is
the identification of the chiral-nematic phase in the absence of any
magnetic order. Although nematic order is algebraically ordered, the
chirality, due to its discrete nature, can undergo a true long range
ordering.

The case for chiral-nematic order at large $J_2 / J_1$ ratio can be
made clearly if we consider the $J_1 - J_2$ model with the discrete
angles $\theta_i = 2\pi n_i /p$, $p=6$, and $n_i$ an integer between
$1$ and $6$. The bi-quadratic $J_2$-interaction turns into a
$3$-state planar model which is known to have a second-order
transition (not KT transition) into an ordered phase\cite{jose}. In
our language, this is the paramagnetic-to-nematic transition. As the
small $J_1$ interaction is introduced, the six-fold spin model within
the nematically ordered phase is governed by the effective
interaction

\be - (J_1 /2) \sum_{\ij} \sigma_i \sigma_j , ~~\sigma_i = \pm 1.
\label{eq:effective-H} \ee
where the Ising variable $\sigma_i$ denotes the two opposite
orientations of the spin. Due to this residual interaction there will
be an Ising phase transition at a temperature $T_I \approx 3.641
\times (J_1/2) \approx 1.82 (1-x)$ according to known results of the
Ising model in two-dimensional triangular lattice. The linear
decrease of $T_I$ with $x$ expected from the effective interaction as
well as the absolute values of the critical temperatures are
consistent with the phase diagram, Fig. \ref{fig:phase-diagram}.
Within the nematic-ordered phase, there are eight spin configurations
allowed for a triangle as shown in Fig. \ref{fig:triangle}. The
chirality for each configuration reads

\be \chi^\triangle_{ijk} =( \sigma_i \sigma_j + \sigma_j \sigma_k +
\sigma_k \sigma_i )/3 , \ee
using the Ising variables. For the downward triangle, the chirality
is the opposite: $\chi^\nabla_{ijk} = - ( \sigma_i \sigma_j +
\sigma_j \sigma_k + \sigma_k \sigma_i )/3$. Then, the net staggered
chirality is given by

\be \chi \sim  \sum (\sigma_i \sigma_j + \sigma_j \sigma_k +
\sigma_k \sigma_i ) \sim \sum_{\ij} \sigma_i \sigma_j . \ee
The final expression, being proportional to the energy, is positive
at any temperature $T$ for a ferromagnetic Ising model given in Eq.
(\ref{eq:effective-H}). Therefore the chirality remains non-zero at
temperature above $T_I$ where magnetic order is lost, but the nematic
order is long-ranged. It is possible that the chirality, due to its
discrete nature, can survive the continuum limit $p\rightarrow
\infty$ and remain long-ranged even as the underlying nematic
correlation is algebraically decaying.

\acknowledgments We thank Gun Sang Jeon, Beom Jun Kim, Dung-Hai Lee
for fruitful discussion. N. N. is supported by Grant-in-Aids under
the grant numbers 16076205, 17105002, 19019004, and 19048015 from the
Ministry of Education, Culture, Sports, Science and Technology of
Japan. S.O. thanks a support from a Grant-in-Aid for Scientific
Research under No. 20046016 from the MEXT of Japan. H. J. H. thanks
the organizers of the workshop, ``Topological Aspects of Solid State
Physics", for support during the final stage of this work.


\begin{thebibliography}{99}

\bibitem{Villain} J. Villain, J. Phys. C.: Solid State Phys. \textbf{10}, 4793 (1977).
\bibitem{MiyashitaShiba84} S. Miyashita and H. Shiba, J. Phys. Soc. Jpn. \textbf{53}, 1145 (1984).
\bibitem{KawamuraTanemura87} H. Kawamura and M. Tanemura, Phys. Rev. B \textbf{36}, 7177
(1987); H. Kawamura, J. Phys.: Condens. Matter \textbf{10}, 4707
(1998).
\bibitem{Baskaran89} G. Baskaran, Phys. Rev. Lett. \textbf{63}, 2524 (1989).
\bibitem{WenWilczekZee89} X. G. Wen, F. Wilczek, and A. Zee, Phys. Rev. B \textbf{39}, 11413 (1989).
\bibitem{Richter92} J. Richter, Phys. Rev. B \textbf{47}, 5794 (1993).
\bibitem{Chalker92} J. T. Chalker, P. C. W. Holdsworth, and E. F. Shender, Phys. Rev. Lett. \textbf{68}, 855 (1992).
\bibitem{KawamuraLi01} H. Kawamura and M. S. Li, Phys. Rev. Lett. \textbf{87}, 187204
(2001).
\bibitem{OnodaNagaosa07} S. Onoda and N. Nagaosa, Phys. Rev. Lett. \textbf{99}, 027206 (2007);
F. David and T. Jolicoeur, Phys. Rev. Lett. \textbf{76}, 3148 (1996).
\bibitem{FSSO08} S. Furukawa \textit{et al.}, ArXiv:0802.3256.
\bibitem{mf} H. Katsura, N. Nagaosa, and A. V. Balatsky, Phys. Rev. Lett. \textbf{95}, 057205 (2005);
C. Jia, S. Onoda, N. Nagosa, and J. H. Han, Phys. Rev. B \textbf{74},
224444 (2006); Phys. Rev. B \textbf{76}, 023708 (2007).
\bibitem{Cinti08} F. Cinti \textit{et al.}, Phys. Rev. Lett. \textbf{100}, 057203 (2008).
\bibitem{olsson} P. Olsson, Phys. Rev. Lett. \textbf{75}, 2758 (1995).
\bibitem{leelee} S. Lee and K.-C. Lee, Phys. Rev. B \textbf{57}, 8472
(1998).
\bibitem{new-literature} S. E. Korshunov, Phys. Rev. Lett.
\textbf{88}, 167007 (2002); M. Hasenbusch, A. Pelissetto and E.
Vicari, J. Stat. Mech.: Theory Exp. (2005) P12002; P. Minnhagen, Beom
Jun Kim, S. Bernhardsson, and G. Cristofano, Phys. Rev. B
\textbf{76}, 224403 (2007).
\bibitem{dhlee} D. H. Lee and G. Grinstein, Phys. Rev.
Lett. \textbf{55}, 541 (1985).
\bibitem{chalker} D. B. Carpenter and J. T. Chalker, J.
Phys. Condens. Mat. \textbf{1}, 4907 (1989).
\bibitem{weber} H. Weber and P. Minnhagen, Phys. Rev. B \textbf{37}, 5986 (1988).
\bibitem{jose} Jorge V. Jos\'{e} et al., Phys. Rev. B \textbf{16},
1217 (1977).

\end{thebibliography}
\end{document}